\documentclass[aps,prl,twocolumn,showpacs,superscriptaddress]{revtex4}  
\usepackage{graphicx} 
\usepackage{dcolumn}  
\usepackage{bm}       
\usepackage{amssymb}  

\def\missET {{\not\!\! E_T}}

\begin{document}


\title{Top Quark Mass Measurement from Dilepton Events at CDF II}

\affiliation{Institute of Physics, Academia Sinica, Taipei, Taiwan 11529, Republic of China} 
\affiliation{Argonne National Laboratory, Argonne, Illinois 60439} 
\affiliation{Institut de Fisica d'Altes Energies, Universitat Autonoma de Barcelona, E-08193, Bellaterra (Barcelona), Spain} 
\affiliation{Baylor University, Waco, Texas  76798} 
\affiliation{Istituto Nazionale di Fisica Nucleare, University of Bologna, I-40127 Bologna, Italy} 
\affiliation{Brandeis University, Waltham, Massachusetts 02254} 
\affiliation{University of California, Davis, Davis, California  95616} 
\affiliation{University of California, Los Angeles, Los Angeles, California  90024} 
\affiliation{University of California, San Diego, La Jolla, California  92093} 
\affiliation{University of California, Santa Barbara, Santa Barbara, California 93106} 
\affiliation{Instituto de Fisica de Cantabria, CSIC-University of Cantabria, 39005 Santander, Spain} 
\affiliation{Carnegie Mellon University, Pittsburgh, PA  15213} 
\affiliation{Enrico Fermi Institute, University of Chicago, Chicago, Illinois 60637} 
\affiliation{Joint Institute for Nuclear Research, RU-141980 Dubna, Russia} 
\affiliation{Duke University, Durham, North Carolina  27708} 
\affiliation{Fermi National Accelerator Laboratory, Batavia, Illinois 60510} 
\affiliation{University of Florida, Gainesville, Florida  32611} 
\affiliation{Laboratori Nazionali di Frascati, Istituto Nazionale di Fisica Nucleare, I-00044 Frascati, Italy} 
\affiliation{University of Geneva, CH-1211 Geneva 4, Switzerland} 
\affiliation{Glasgow University, Glasgow G12 8QQ, United Kingdom} 
\affiliation{Harvard University, Cambridge, Massachusetts 02138} 
\affiliation{Division of High Energy Physics, Department of Physics, University of Helsinki and Helsinki Institute of Physics, FIN-00014, Helsinki, Finland} 
\affiliation{University of Illinois, Urbana, Illinois 61801} 
\affiliation{The Johns Hopkins University, Baltimore, Maryland 21218} 
\affiliation{Institut f\"{u}r Experimentelle Kernphysik, Universit\"{a}t Karlsruhe, 76128 Karlsruhe, Germany} 
\affiliation{High Energy Accelerator Research Organization (KEK), Tsukuba, Ibaraki 305, Japan} 
\affiliation{Center for High Energy Physics: Kyungpook National University, Taegu 702-701; Seoul National University, Seoul 151-742; and SungKyunKwan University, Suwon 440-746; Korea} 
\affiliation{Ernest Orlando Lawrence Berkeley National Laboratory, Berkeley, California 94720} 
\affiliation{University of Liverpool, Liverpool L69 7ZE, United Kingdom} 
\affiliation{University College London, London WC1E 6BT, United Kingdom} 
\affiliation{Massachusetts Institute of Technology, Cambridge, Massachusetts  02139} 
\affiliation{Institute of Particle Physics: McGill University, Montr\'{e}al, Canada H3A~2T8; and University of Toronto, Toronto, Canada M5S~1A7} 
\affiliation{University of Michigan, Ann Arbor, Michigan 48109} 
\affiliation{Michigan State University, East Lansing, Michigan  48824} 
\affiliation{Institution for Theoretical and Experimental Physics, ITEP, Moscow 117259, Russia} 
\affiliation{University of New Mexico, Albuquerque, New Mexico 87131} 
\affiliation{Northwestern University, Evanston, Illinois  60208} 
\affiliation{The Ohio State University, Columbus, Ohio  43210} 
\affiliation{Okayama University, Okayama 700-8530, Japan} 
\affiliation{Osaka City University, Osaka 588, Japan} 
\affiliation{University of Oxford, Oxford OX1 3RH, United Kingdom} 
\affiliation{University of Padova, Istituto Nazionale di Fisica Nucleare, Sezione di Padova-Trento, I-35131 Padova, Italy} 
\affiliation{University of Pennsylvania, Philadelphia, Pennsylvania 19104} 
\affiliation{Istituto Nazionale di Fisica Nucleare Pisa, Universities of Pisa, Siena and Scuola Normale Superiore, I-56127 Pisa, Italy} 
\affiliation{University of Pittsburgh, Pittsburgh, Pennsylvania 15260} 
\affiliation{Purdue University, West Lafayette, Indiana 47907} 
\affiliation{University of Rochester, Rochester, New York 14627} 
\affiliation{The Rockefeller University, New York, New York 10021} 
\affiliation{Istituto Nazionale di Fisica Nucleare, Sezione di Roma 1, University of Rome ``La Sapienza," I-00185 Roma, Italy} 
\affiliation{Rutgers University, Piscataway, New Jersey 08855} 
\affiliation{Texas A\&M University, College Station, Texas 77843} 
\affiliation{Istituto Nazionale di Fisica Nucleare, University of Trieste/\ Udine, Italy} 
\affiliation{University of Tsukuba, Tsukuba, Ibaraki 305, Japan} 
\affiliation{Tufts University, Medford, Massachusetts 02155} 
\affiliation{Waseda University, Tokyo 169, Japan} 
\affiliation{Wayne State University, Detroit, Michigan  48201} 
\affiliation{University of Wisconsin, Madison, Wisconsin 53706} 
\affiliation{Yale University, New Haven, Connecticut 06520} 
\author{A.~Abulencia}
\affiliation{University of Illinois, Urbana, Illinois 61801}
\author{D.~Acosta}
\affiliation{University of Florida, Gainesville, Florida  32611}
\author{J.~Adelman}
\affiliation{Enrico Fermi Institute, University of Chicago, Chicago, Illinois 60637}
\author{T.~Affolder}
\affiliation{University of California, Santa Barbara, Santa Barbara, California 93106}
\author{T.~Akimoto}
\affiliation{University of Tsukuba, Tsukuba, Ibaraki 305, Japan}
\author{M.G.~Albrow}
\affiliation{Fermi National Accelerator Laboratory, Batavia, Illinois 60510}
\author{D.~Ambrose}
\affiliation{Fermi National Accelerator Laboratory, Batavia, Illinois 60510}
\author{S.~Amerio}
\affiliation{University of Padova, Istituto Nazionale di Fisica Nucleare, Sezione di Padova-Trento, I-35131 Padova, Italy}
\author{D.~Amidei}
\affiliation{University of Michigan, Ann Arbor, Michigan 48109}
\author{A.~Anastassov}
\affiliation{Rutgers University, Piscataway, New Jersey 08855}
% ...
\collaboration{CDF Collaboration}
\noaffiliation


\begin{abstract}
We report a measurement of the top quark mass using events collected by the CDF II Detector from $p\overline{p}$ collisions at $\sqrt s = 1.96$ TeV at the Fermilab Tevatron.  We calculate a likelihood function for the top quark mass in events that are consistent with $t\overline{t} \to \overline{b}\ell^{-}\overline{\nu}_{\ell}b\ell'^{+}\nu'_\ell$ decays.  The likelihood is formed as the convolution of the leading-order matrix element and detector resolution functions. The joint likelihood is the product of likelihoods for each of 33 events collected in 340 pb$^{-1}$ of integrated luminosity, yielding a top quark mass $M_{t} = 165.2 \pm 6.1(\textrm{stat.}) \pm 3.4(\textrm{syst.}) \mathrm{~GeV}/c^2$.  This first application of a matrix-element technique to $t\overline{t} \to b\ell^{+}\nu_\ell\overline{b}\ell'^{-}\overline{\nu}_{\ell'}$ decays gives the most precise single measurement of $M_t$ in dilepton events.  Combined with other CDF Run II measurements using dilepton events, we measure $M_{t} = 167.9 \pm 5.2 (\textrm{stat.})  \pm 3.7 (\textrm{syst.}) \mathrm{~GeV}/c^2$.
\end{abstract}


\maketitle 

\date{\today}

Precision measurements of the top quark mass, $M_t$, place constraints on the masses of particles to which the top quark contributes radiative corrections, including the unobserved Higgs boson~\cite{higgs} and particles in extensions to the standard model~\cite{topconstraints}. At the Tevatron, top quarks are primarily produced in pairs. The dilepton channel, consisting of the decays $t\overline{t} \rightarrow \overline{b}\ell^{-}\overline{\nu}_{\ell}b\ell'^{+}\nu'_\ell$, has a small branching fraction but allows measurements which are less reliant on the calibration of the jet energy scale, the dominant systematic uncertainty, than measurements in channels with hadronic $W$ decays.  A discrepancy from other channels could indicate contributions from new processes~\cite{newsources}. 

The reconstruction of the top quark mass from dilepton events poses a particular challenge as the two neutrinos from $W$ decays are undetected.  Previous measurements in this channel~\cite{d0dilep,cdfdilep} using Run I data have calculated the mass by making several kinematic assumptions and integrating over the remaining unconstrained quantity.  To extract maximum information from the small sample of dilepton events, we adapt a technique pioneered for the analysis of $t\overline{t} \to b\ell\nu_\ell\overline{b}q\overline{q}'$ decays~\cite{dg, kondo, estrada, d0me,canelli}. This technique uses the leading-order production cross-section and a parameterized description of the jet energy resolution.  Making minimal kinematic assumptions and integrating over six unconstrained quantities, we obtain per-event likelihoods in top quark mass which can be directly multiplied to obtain the joint likelihood from which $M_t$ is determined.

This Letter reports a measurement with data collected by the CDF II detector between March 2002 and August 2004, with an integrated luminosity of 340~pb$^{-1}$, yielding the most precise single measurement of $M_t$ in dilepton events, with statistical uncertainty approximately half that of Run I measurements~\cite{d0dilep,cdfdilep}. We also report a combination of this measurement with other recent measurements using CDF Run II data in the dilepton channel.

The CDF II detector~\cite{CDF,coordinates} is an azimuthally and forward-backward symmetric apparatus designed to study $p\overline{p}$ collisions at the Fermilab Tevatron. It consists of a magnetic spectrometer surrounded by calorimeters and muon chambers. The charged particle tracking system is immersed in a 1.4~T magnetic field parallel to the $p$ and $\overline{p}$ beams. Calorimeters segmented in $\eta$ and $\phi$ surround the tracking system and measure the energies of interacting particles.  The electromagnetic and hadronic calorimeters are lead-scintillator and iron-scintillator sampling devices. Drift chambers located outside the central hadron calorimeters detect muons. 

The data are collected with lepton triggers that require events to have an electron or muon with $p_T>18$~GeV$/c$.  After offline reconstruction, we select events with (i) two leptons, each with $p_T > 20$~GeV$/c$, (ii) significant missing energy transverse to the beam direction ($\missET$), and (iii) two jets, each with $E_T > 15$ GeV. The selection is defined as ``DIL'' in~\cite{dilepton} and was used to measure the cross section in the dilepton channel. These requirements yield 33 events in the sample reported in this Letter.

The probability  density for $t\overline{t}$ decays is expressed as $P_s(\mathbf{x} | M_t)$, where $M_t$ is the top quark pole mass and $\mathbf{x}$ contains the measured lepton and jet momenta.  We calculate $P_s(\mathbf{x} | M_t)$ using the theoretical description of the $t\overline{t}$ production process expressed with respect to $\mathbf{x}$, $P_s(\mathbf{x} | M_t) = \frac{1}{\sigma(M_t)} \frac{ d\sigma(M_t)}{d\mathbf{x}} $, where $\frac{ d\sigma}{d\mathbf{x}}$ is the differential cross section and $\sigma$ is the total cross section.

To evaluate the probability density, we integrate the leading-order matrix element over quantities which are not directly measured by the detector, \emph{i.e.}, neutrino  and quark energies. We assume that  lepton momenta are perfectly measured, that quark angles are perfectly measured by the corresponding jets, and that the two most energetic jets correspond to the $b$ quarks from top quark decay.   Quark energies, while not directly measured, are estimated from the observed energies of the corresponding jets. We define the transfer function $W(p,j)$ to be the probability of measuring jet energy $j$ given quark energy $p$. We approximate $W(p,j)$ as a sum of two Gaussians fitted to the predicted distribution of quark-jet energy difference from $t\overline{t}$ events generated with {\sc Herwig}~\cite{Herwig} and the CDF II detector simulation~\cite{cdfsim}.  The expression for the probability density at a given mass for a specific event can be written as

\begin{eqnarray}
P_s(\mathbf{x}|M_t) = \frac{1}{\sigma(M_t)}\int d\Phi |{\mathcal M}_{t\overline{t}}(q_i,p_i;M_t)|^2 \\
\times \prod_{jets} W(p_i,j_i) f_{PDF}(q_1)f_{PDF}(q_2), \nonumber
\label{eqn:sigprob}
\end{eqnarray}
\noindent
where the integral is over the momenta of the initial and final state particles, $q_1$ and $q_2$ are the incoming momenta, $p_i$ are the outgoing momenta, $f_{PDF}(q_i)$ are the parton distribution functions~\cite{cteq} and ${\mathcal M}_{t\overline{t}}(q_i,p_i;M_t)$ is the $t\overline{t}$ production and decay matrix element as defined in~\cite{meref1, meref2} for the process $q\overline{q} \rightarrow t\overline{t} \rightarrow b\ell^{+}\nu_\ell\overline{b}\ell'^{-}\overline{\nu}_{\ell'}$.  While up to 15\% of $t\overline{t}$ pairs at the Tevatron are produced by gluon-gluon fusion ($gg\to t\overline{t}$), this term can be excluded from the matrix element with negligible effect on the measurement.   The term, $1/\sigma(M_t)$, in front of the integral ensures that the normalization condition for the probability density, $\int d\mathbf{x}\ P_s(\mathbf{x}|M_t) = 1$, is satisfied.  The effect of assumptions and approximations are measured in Monte Carlo experiments of fully realistic events, described below. Where needed, a correction is taken. 

We calculate the probability for the dominant background processes, $P_{bg}(\mathbf{x})$, and form the generalized per-event probability density $P(\mathbf{x}|M_t) = P_{s}(\mathbf{x}|M_t) p_s(M_t) + P_{bg_1}(\mathbf{x}) p_{bg_1} + P_{bg_2}(\mathbf{x}) p_{bg_2}\cdots$, where $p_s(M_t)$ and $p_{bg_i}$ are determined from the expected fractions of signal and background events (see Table~\ref{tab:bkg}).  The $P_{bg}$ are calculated in analogy to Equation 1, with the background matrix element evaluated numerically using algorithms adapted from the {\sc Alpgen}~\cite{alpgen} generator.  We calculate the probabilities for backgrounds arising from $Z/\gamma^* \to ee,\mu\mu$ plus associated jets, W+ $\ge 3$ jets where one jet is incorrectly identified as a lepton, and $WW$ plus associated jets.  Smaller backgrounds, comprising 11\% of the expected background, are not modeled.  Application of the background probabilities reduces the expected statistical uncertainty by 15\%. 

\begin{table}
\caption {Expected numbers of signal and background events for a data sample of integrated luminosity of 340~pb$^{-1}$. Other backgrounds are negligible; all numbers have an additional correlated 6\% error from uncertainty in the sample luminosity. }

\begin{ruledtabular}
\begin{tabular}{lr}
Source & Events \\
\hline
 Expected $t\overline{t}$~($M_t=178$~GeV/$c^2$, $\sigma=6.1$~pb) &$15.9\pm 1.4$ \\
 Expected Background & $10.5\pm 1.9$ \\
\ \ \  Drell-Yan ($Z/\gamma^*$) & $5.5\pm 1.3$\ \ \ \ \ \ \ \ \ \ \\
\ \ \  Misidentified Lepton & $3.5\pm 1.4$\ \ \ \ \ \ \ \ \ \ \\
\ \ \  Diboson ($WW/WZ$) & $1.6\pm 0.2$\ \ \ \ \ \ \ \ \ \ \\
 \hline
 Total Expected & $26.4\pm 2.3$ \\
 Run II Data  & 33 \\
\end{tabular}

\end{ruledtabular}
\label{tab:bkg}\rm
\end{table}%

\begin{figure}[!htbp]
\begin{center}
\includegraphics[width=2.8in]{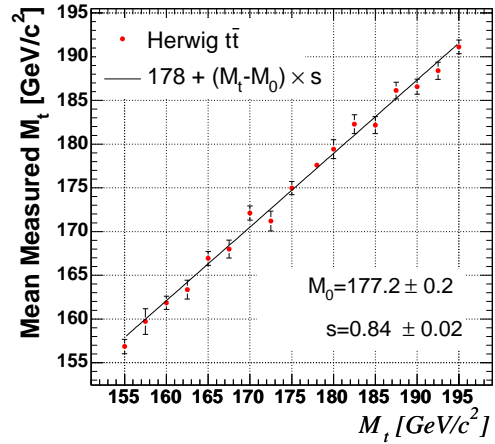}
\end{center}
\caption{Mean measured $M_t$ in Monte Carlo experiments of signal and background events at varying top quark mass. The solid line is a linear fit to the points. }
\label{fig:response}
\end{figure}

The posterior probability density in top quark mass is the product of a flat Bayesian prior and the joint likelihood, a product of the individual event likelihoods. The mass measurement ($M_t$) is the mean of the posterior probability, and the statistical uncertainty ($\Delta M_t$) is the standard deviation.  We calibrate the method using Monte Carlo experiments of signal and background events. Signal events are generated with {\sc Herwig} for top quark masses from 155~GeV$/c^2$ to 195~GeV$/c^2$. Misidentified leptons are modeled using events from the data, while all other backgrounds are generated with {\sc Alpgen} ($Z/\gamma^*\to ee,\mu\mu$) or {\sc Pythia}~\cite{pythia} ($Z/\gamma^*\to \tau\tau, WW,WZ,ZZ$) Monte Carlo. The numbers of signal and background events in each pseudo-experiment are chosen according to Poisson distributions with mean values given in Table~\ref{tab:bkg}. The estimates for $t\overline{t}$ yields take into account the mass dependence of the cross section~\cite{ttbarcross} and acceptance. The response of the method for Monte Carlo experiments with both signal and background is shown in Fig.~\ref{fig:response}. The response is consistent with a linear dependence on top quark mass, but has a slope that is less than unity due to the incomplete modeling of background contributions.   Corrections, ${ M_t \rightarrow 177.2 + (M_t-178.0)/0.84, \Delta M_t \rightarrow \Delta M_t/0.84 }$ are derived from this response and applied to both $M_t$ and $\Delta M_t$ measured in data.

Examining the width of the pull distributions in these Monte Carlo experiments, we find that the statistical uncertainty is underestimated by a factor of 1.51, independent of top quark mass. This results from simplifying assumptions described above, made to ensure the computational tractability of the integrals in Equation 1.  The largest effects are from jets which originate from initial or final state radiation rather than $b$-quark hadronization, imperfect lepton momentum resolution, imperfect jet angle resolution, and unmodeled backgrounds.  Correcting by this factor of 1.51, we estimate the mean statistical uncertainty to be 9.4$\mathrm{~GeV}/c^2$ if $M_t=178 \mathrm{~GeV}/c^2$~\cite{runone} or $7.8 \mathrm{~GeV}/c^2$ if $M_t=165 \mathrm{~GeV}/c^2$.

\begin{figure}[!h]
\begin{center}
\includegraphics[width=2.8in]{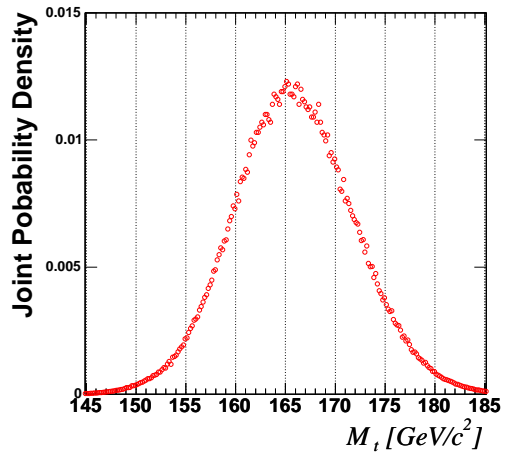}
\end{center}
\caption{Joint posterior probability density as a function of the top quark mass for the 33 candidate events in data, after all corrections. Systematic uncertainties are not shown.} 
\label{fig:dataprob}
\end{figure}

We apply the method described in this Letter to the 33 candidate events observed in the data. Including all corrections described above, we measure $M_{t}=165.2\pm 6.1 (\textrm{stat.}) \textrm{ GeV}/c^2$. Figure ~\ref{fig:dataprob} shows the the joint probability density, without systematic uncertainty, for the events in our data set.

The measured statistical uncertainty is consistent with the distribution of statistical uncertainties in Monte Carlo experiments where signal events with $M_t = 165\textrm{ GeV}/c^2$ are chosen according to a Poisson distribution with mean $N_{t\overline{t}} = 21.7$ events. This number of events corresponds to the cross section and acceptance at $M_t = 165\textrm{ GeV}/c^2$. Of these Monte Carlo experiments, 17\% yielded a statistical uncertainty less than 6.1 GeV$/c^2$.

\begin{table}
\caption{Summary of systematic uncertainties.}
\begin{ruledtabular}
\begin{tabular}{lr}

Source & $\Delta M_t$ (GeV/$c^2$)\\\hline
Jet energy scale & 2.6\\
Limited background statistics & 1.2\\
PDFs & 1.0\\
Generator & 0.8\\
Background modeling & 0.8\\
FSR modeling & 0.7\\
ISR modeling & 0.5\\
Response correction & 0.4\\
Sample composition uncertainty & 0.3 \\
\hline
Total &  3.4\\
\end{tabular}
\end{ruledtabular}
\label{tab:sys} \rm
\end{table}

There are several sources of systematic uncertainty in our measurement which are summarized in Table~\ref{tab:sys}. 
The largest is due to the uncertainty in the jet energy scale~\cite{jes}, which we estimate at 2.6~GeV/$c^2$ by varying the scale within its uncertainty.   An uncertainty of 1.2 GeV$/c^2$ comes from the limited number of background events available for Monte Carlo experiments. The contribution from uncertainties in the PDFs are estimated by using different PDF sets (CTEQ5L~\cite{cteq} vs. MRST72~\cite{mrst}), different values of $\Lambda_{QCD}$, varying the eigenvectors of the CTEQ6M~\cite{cteq} set, and varying the initial state contributions of $gg$ and $q\overline{q}$; the total corresponding uncertainty added in quadrature is 1.0~GeV$/c^2$.  Dependence on the Monte Carlo generator is estimated as the difference in the extracted top quark mass from {\sc Pythia} events and {\sc Herwig} events; this amounts to 0.8~GeV$/c^2$. We observe no difference in the extracted top quark mass in events from a leading-order and a next-to-leading order generator~\cite{nlo1,nlo2}. We estimate the uncertainty coming from  modeling of the two largest sources of background, $Z/\gamma^*$ and events with a misindentified lepton, to be 0.8~GeV$/c^2$. The uncertainty due to imperfect modeling of initial state (ISR) and final state (FSR) QCD radiation is estimated by varying the amount of ISR and FSR in simulated events~\cite{ljetsprd} and is measured to be 0.7~GeV$/c^2$ for FSR and 0.5~GeV$/c^2$ for ISR.   The uncertainty in the mass due to uncertainties in the response correction shown in Fig.~\ref{fig:response} is 0.4~GeV$/c^2$.  The contribution from uncertainties in background composition is estimated by varying the background estimates from Table~\ref{tab:bkg} within their uncertainties and amounts to 0.3 GeV$/c^2$. Adding all of these contributions together in quadrature yields a total systematic uncertainty of 3.4 GeV/$c^2$.

Applications of other dilepton techniques~\cite{cdfdilep2prd} using the same CDF Run II data yield values consistent with this result. The four results are combined using a standard method~\cite{blue}. We determine statistical correlations between the measurements using Monte Carlo experiments, and assume systematic uncertainties to be 100\% correlated except the few that are method-specific which are assumed to be uncorrelated. Table~\ref{tab:dilepresults} gives the four dilepton measurements, their statistical correlations, and their weight in the combination. The NWA method uses looser event selection and thus has a smaller statistical correlation. Correlations significantly less than unity suggest that each method extracts unique information.

\begin{table}[htdp]
\caption{Measurements of $M_t$ in the dilepton channel with statistical and systematic errors, their statistical correlations and the weight of each measurement in the combined result.}
\begin{ruledtabular}
\begin{tabular}{llccccr}
Method & Result (GeV$/c^2$) &\multicolumn{4}{c}{Correlation Matrix} & Weight \\
\hline
This Letter & $165.2^{+6.1}_{-6.1} \pm 3.4$ & 1 & & & & 0.47 \\
NWA~\cite{cdfdilep2prd} &$170.7^{+6.9}_{-6.5}\pm 4.6$& 0.12 & 1 & & & 0.36 \\
KIN~\cite{cdfdilep2prd} &$169.5^{+7.7}_{-7.2} \pm 4.0$& 0.40 & 0.14 & 1  & & 0.18 \\
PHI~\cite{cdfdilep2prd} &$169.7^{+8.9}_{-9.0} \pm 4.0$& 0.43 & 0.25 & 0.35 & 1 & 0.00 \\
\end{tabular}
\end{ruledtabular}
\label{tab:dilepresults}
\end{table}

Combining the four CDF Run II dilepton measurements we obtain 

\begin{displaymath}
M_{t}=167.9\pm 5.2 (\textrm{stat.})\pm 3.7 (\textrm{syst.})  \textrm{ GeV}/c^2.
\end{displaymath}
\noindent
This combined result is consistent with the current world average~\cite{worldave} and the single most precise measurement in the lepton+jets channel~\cite{ljetsprl}. 

We thank the Fermilab staff and the technical staffs of the participating institutions for their vital contributions. This work was supported by the U.S. Department of Energy and National Science Foundation; the Italian Istituto Nazionale di Fisica Nucleare; the Ministry of Education, Culture, Sports, Science and Technology of Japan; the Natural Sciences and Engineering Research Council of Canada; the National Science Council of the Republic of China; the Swiss National Science Foundation; the A.P. Sloan Foundation; the Bundesministerium f\"ur Bildung und Forschung, Germany; the Korean Science and Engineering Foundation and the Korean Research Foundation; the Particle Physics and Astronomy Research Council and the Royal Society, UK; the Russian Foundation for Basic Research; the Comisi\'on Interministerial de Ciencia y Tecnolog\'{\i}a, Spain; in part by the European Community's Human Potential Programme under contract HPRN-CT-2002-00292; and the Academy of Finland.


\begin{thebibliography}{99}
\bibitem{coordinates}
CDF uses a cylindrical coordinate system with the $z$ axis along the proton beam axis. Pseudorapidity is $\eta \equiv -\ln(\tan (\theta/2))$, where $\theta$ is the polar angle, and $\phi$ is the azimuthal angle relative to the proton beam direction, while $p_T = |p|\ sin(\theta)$, $E_T = E\ sin(\theta)$.
\end{thebibliography}
\end{document}